\documentclass[12pt, letterpaper]{article}
\usepackage[top=1in, bottom=1.5in, left=1in, right=1in]{geometry}
\usepackage[utf8]{inputenc}
\usepackage{booktabs}
\usepackage{hyperref}
\usepackage{natbib}
\usepackage{amsmath}
\usepackage{amssymb}
\usepackage{authblk}

\title{Continuous Cadence Acquisition of the LSST \\ Deep Drilling Fields}
\author[ ]{Keaton J.\ Bell$^1$ and J.\ J.\ Hermes$^2$, \\ with the support of the LSST Transient and Variable Stars Collaboration} 
\affil[1]{Max Planck Institute for Solar System Research}
\affil[2]{University of North Carolina at Chapel Hill}

\date{November 2018}

\begin{document}

\maketitle

\begin{abstract}
To extend LSST's coverage of the transient and variable sky down to minute timescales, we propose that observations of the Deep Drilling Fields are acquired in sequences of continuous exposures each lasting 2--4 hours. This will allow LSST to resolve rapid stellar variability such as short-period pulsations, exoplanet transits, ultracompact binary systems, and flare morphologies, while still achieving the desired co-added depths for the selected fields. The greater number of observations of each Deep Drilling Field pushes these mini-surveys deep in terms of both sensitivity to low-amplitude variability and co-added depth. Saving the individual 15-second exposures will yield an effective Nyquist limit of $\approx0.031$\,Hz (32\,seconds). Resolved short-period variability of targets in these fields will aid the interpretation of sparse observations of a greater number of variables in the main survey. If this cadence strategy conflicts with the science goals of individual Deep Drilling Fields, at least a subset of the additional observations of each field should be obtained continuously.  This strategy should also be considered for the proposed Galactic Plane mini survey, which will observe a greater number of stellar variables and transients.

\end{abstract}

\section{White Paper Information}
Correspondence can be directed to Keaton Bell: \href{mailto:bell@mps.mpg.de}{bell@mps.mpg.de} 
\begin{enumerate} 
\item {\bf Science Category:} Exploring the Transient Optical Sky

\item {\bf Survey Type Category:} Deep Drilling fields, mini survey

\item {\bf Observing Strategy Category:} a specific observing strategy to enable specific time domain science, 
	that is relatively agnostic to where the telescope is pointed
\end{enumerate}

\clearpage

\section{Scientific Motivation}

In its main Wide-Fast-Deep survey, LSST will achieve a thorough census of the variable sky on timescales exceeding the minimum typical revisit time of $\sim1$\,hr.  The frequency of revisits in the main survey is limited to achieve LSST's broad science goals.  However, LSST will observe each of a few ``Deep Drilling Fields'' (DDFs) $\approx 5$ more times than the typical 825 visits for the 10-year main survey.  In addition to achieving higher co-added survey depths, the additional observations will be able to better sample time domain variability in these specific fields.  While some of the DDF locations and filter distributions have been discussed or selected, the detailed drilling cadences have not been established.  We propose that acquiring the additional observations of the DDFs in continuous, multi-hour sequences will extend the completeness of LSST's survey of the transient and optical sky down to $\sim$minute timescales while decreasing observational overhead of the DDF mini surveys due to less slewing.

There are many interesting astrophysical processes that are known to cause photometric variations on $\sim$minute timescales. For example, some classes of pulsating stars, including pulsating white dwarfs \citep[e.g.,][]{Mukadam2006}, pulsating hot subdwarfs \citep[sdBs;][]{Kilkenny2007}, blue large-amplitude pulsators \citep[BLAPs;][]{Pietrukowicz2017}, and rapidly oscillating Ap stars \citep{Kurtz1990}, have typical periods of $\sim10$\,minutes.
These variations will be detectable given LSST's photometric precision, but severe undersampling of the Wide-Fast-Deep survey will hamper accurate period determinations that would be useful for asteroseismology or other in-depth analyses of individual objects.  If the regular revisit times are more frequent than twice per intrinsic period, these signals will not be obscured by the effects of aliasing.

Many transients also have much shorter durations than typical LSST revisits, and these will only be observed at most once per event by the main survey.  For example, most flares from M dwarf stars have durations shorter than an hour \citep[e.g.,][]{Moffett1974}, and deep transits of planetary debris or even intact planets around white dwarfs will only last a couple of minutes \citep{Cortes2018,Lund2018}.  The continuous cadence proposed here will enable the shapes of such events in the light curves to be recorded and studied in detail.  The examples that are recorded at high cadence in the DDFs will serve as guides for interpreting similar transients of objects that are only detected in sparse, individual visits by the main survey.

With the DDF fields receiving $\approx5\times$ more visits than the main survey, the DDF mini-surveys will be sensitive to lower-amplitude photometric variability.  With $N$ total visits per field, the noise level in a periodogram and the uncertainty on frequency, amplitude, and phase from least-squares fits to the time series all scale as $\propto 1/\sqrt{N}$ \citep[e.g.,][]{Montgomery1999}.  These will all be better by more than a factor of 2 for the DDFs compared to the main survey, making these mini-surveys ``deep'' not only in co-added depth, but also variability amplitude.

While we advocate for continuous visits lasting at least 2\,hr to fill the gap below the typical main-survey revisit timescale, longer visits are preferred to improve frequency resolution of the data.  In all practicality, for individual observations spaced by $\Delta t$ acquired continuously for sequences of duration $T$, the effective limiting upper Nyquist frequency of observations is $1/(2\Delta t)$, with an effective frequency resolution of $1/T$.  This slight simplification ignores the longer total baseline of observations, and while this entire $\approx10$-year baseline technically defines the frequency resolution, in reality the $1/T$ resolution only becomes filled by inaccurate aliases that reach the higher resolution.  Variable sources that have multiple periodicities separated by $<1/T$ will be realistically indecipherable from a single periodicity, and so exacting time domain science---such as probing the interiors of pulsating stars with asteroseismology---will benefit from longer individual visits.

To obtain a high Nyquist frequency, across which higher-frequency intrinsic signals are reflected as inaccurate aliases that are easy to misinterpret, the individual 15-second exposures should be saved and processed for the high-cadence survey.  This will yield a Nyquist limit of 0.03125\,Hz, assuming two exposures take 32 seconds to obtain and read out.

Even when limited to the DDFs, LSST will provide a deeper survey of many short-period variables and transients than previous surveys of this type. Shallower examples that have been motivated by similar science and that demonstrate the value of such observations include the OmegaWhite survey on the VLT \citep{Macfarlane2015,Toma2016} and the ``Fast and Furious'' minisurvey covering the Galactic plane with the Zwicky Transient Facility \citep[ZTF;][]{ZTF}.  These other efforts have particularly focused on discovering ultracompact binaries, the precise period determination of which will be complicated by aliasing from cycle-count-ambiguities if sparsely sampled. While many of the DDFs will target extragalactic fields and our rapid variability science is primarily motivated by stellar astrophysics, the wide field of view of the LSST images ensures that many stars will be resolved in each of these fields, especially faint white dwarfs that can show many types of $\sim$minute timescale variations.  If this proposed cadence is in conflict with specific science criteria of some of the DDFs, application to any subset of the DDFs remains extremely valuable, with preference given to those containing the greatest number of resolved stars.  At the least, we urge that a subset of the additional observations of each DDF be obtained in at least one continuous run of 2--4 hours---even these more limited high-cadence observations will allow us to disentangle some of the severe observational frequency aliases that will result from the otherwise sparse sampling.

Although the various Galactic Plane mini surveys being proposed (see white papers by, e.g., Lund et al.; Street et al.) request fewer observations than the DDFs, the concentration of stellar variables in these fields make them particularly valuable for rapid, continuous monitoring (e.g., the ZTF ``Fast and Furious'' mini survey targets the Galactic Plane specifically).  While this white paper proposes a specific cadence rather than survey fields, we submit that continuous acquisition of the observations requested by Galactic Plane mini survey proposals should also be considered.

\clearpage

\section{Technical Description}

\subsection{High-level description}

If DDFs receive five times as many exposures as the main survey, the total time exposing each field is $\approx34.4$\,hr. The science requirements for a co-added survey depth in these fields is directly satisfied by the total number of observations, assuming the same average single-visit depth (same average seeing/brightness conditions).  In order to obtain an interpretable record of short-timescale photometric variability in these fields, these observations would ideally be acquired in continuous sequences lasting 2--4 hours each.  This strategy yields a high Nyquist frequency with minimal aliasing.  Saving and processing the data from each of the 15-second exposures will produce a Nyquist limit twice that attained if the images are only stored from $2\times15$-second pairs.

\subsection{Footprint -- pointings, regions and/or constraints}

We propose only a modification of the cadence for the DDFs while allowing the selection of those pointings to be decided by the nearly independent science considerations of which fields benefit most from higher co-added survey depth.  Our proposed drilling cadence could be limited to single extended 2--4 hour runs for DDFs where obtaining all observations in a continuous mode would conflict with other cadence requirements that were integral to the selection of these fields---this would still enable much high-speed variability science in these fields, though with less sensitivity to low-amplitude variations in fainter stars. Our request for high-cadence observations is motivated primarily for sampling stellar variability; if only a subset of DDFs can be observed at continuous cadence, preference should be given to Galactic and LMC/SMC fields, though even extragalactic fields will include a valuable number of foreground stars.  Adopting a continuous cadence for the Galactic Plane mini surveys under consideration would also yield a tremendous record of rapid stellar variability in the Galaxy.

\subsection{Image quality}
Suitable image quality to achieve the desired DDF co-added depths will be sufficient for the proposed cadence.

\subsection{Individual image depth and/or sky brightness}

The sources brighter than the limiting magnitudes of the individual images will be the targets that can be directly analyzed in the high-cadence observations.  It will be especially informative to compare short-timescale variability in different bandpasses. To maximize the number of sources with high-speed coverage in most filters, observations in bandpasses with generally lower total throughput should be preferred in dark sky conditions.  We note that our requested readout of the 15-second exposures will not be individually as deep as the $2\times15$\,s coadds, but they can be combined later.

\subsection{Co-added image depth and/or total number of visits}
We anticipate that the co-added depth will be constrained in the selection of DDFs. A greater number of total observations decreases the noise floor of periodograms, enabling lower-amplitude variability to be detected.

\subsection{Number of visits within a night}

For 2--4 hours of continuous coverage, each sequence within a given night will consist of roughly 450--900 15-second exposures (assuming 1-second overhead for each).  This sequence length provides decent frequency resolution ($\approx$70--140\,$\mu$Hz), but runs as short as 1 hour would provide coverage of the frequency range below the typical Wide-Fast-Deep survey revisit timescales. Longer individual runs are preferred to yield better frequency resolution for studying multi-periodic variables and more precise measurements of $\lesssim1$-hour periods.

\subsection{Distribution of visits over time}

Summing the total time spent exposing and reading out images for each DDF will reveal some number of multi-hour sequences that each can be broken down into.  During a visit, the exposures should be made continuously back-to-back, but we do not define a preferred timescale for spacing the revisits, and these will perhaps be different for each field depending on the reason for its selection.  To reliably obtain the desired sequence lengths, these should be initiated only when the weather forecasts predict multiple hours of observable conditions.

\subsection{Filter choice}


Depending on the science motivation for selecting each field, the distributions of filters may be different. Changing filters occasionally during a sequence of exposures would not significantly decrease the Nyquist frequency and should be done as required by the mini survey design.  \citet{VanderPlas2015} present analysis tools for accounting for observations made in multiple bandpasses when calculating a periodogram.  However, longer segments of observations in individual filters are preferred, and with a limited number of allowed filter changes over the 10-year survey, limiting filter changes here may create a bit more flexibility for additional filter changes in other surveys.

\subsection{Exposure constraints}

Access to the individual 15-second exposures will maximize the effective Nyquist frequency below which intrinsic frequencies of variability can be accurately recovered. 

\subsection{Other constraints}

No other constraints. 

\subsection{Estimated time requirement}

The time required to observe each DDF will mostly be determined by the number of observations required to reach the target co-added depths.  However, this observing strategy will save time on overheads by requiring fewer slews to the DDFs.  If each field is monitored for a total of $\approx39$\,hr in  $10\times\approx4$-hour sequences, the slew overhead will only be $10\times120 = 1200$\,sec, making LSST more efficient overall (compared to 99000\,sec of slewing to each DDF that might otherwise be observed by individual $2\times15$ second visits).

\begin{table}[ht]
    \centering
    \begin{tabular}{l|l|l|l}
        \toprule
        Properties & Importance \hspace{.3in} \\
        \midrule
        Image quality & 2    \\
        Sky brightness & 2 \\
        Individual image depth & 2  \\
        Co-added image depth & 2  \\
        Number of exposures in a visit   & 1  \\
        Number of visits (in a night)  & 1  \\ 
        Total number of visits & 1  \\
        Time between visits (in a night) & 1 \\
        Time between visits (between nights)  & 3  \\
        Long-term gaps between visits & 3 \\
        Other (please add other constraints as needed) & 3 \\
        \bottomrule
    \end{tabular}
    \caption{{\bf Constraint Rankings:} Summary of the relative importance of various survey strategy constraints. Ranked from 1=very important, 2=somewhat important, 3=not important.}
        \label{tab:obs_constraints}
\end{table}

\subsection{Technical trades}
To aid in attempts to combine this proposed survey modification with others, the following questions are addressed:
\begin{enumerate}
    {\it \item  What is the effect of a trade-off between your requested survey footprint (area) and requested co-added depth or number of visits?}
    
    The total number of visits per DDF may be decreased for an increased number of DDFs.  Many hours of continuous coverage, broken up into a few 2--4 hour segments, would be sufficient for studying $\lesssim1$-hour timescale variability. More fields with fewer overall exposures would be preferred for increasing the number of targets with high-speed coverage.
    
    {\it \item If not requesting a specific timing of visits, what is the effect of a trade-off between the uniformity of observations and the frequency of observations in time? e.g. a `rolling cadence' increases the frequency of visits during a short time period at the cost of fewer visits the rest of the time, making the overall sampling less uniform.}
    
    We are requesting a specific timing of visits, with uniform-cadence, high-frequency observations.
    
    {\it \item What is the effect of a trade-off on the exposure time and number of visits (e.g. increasing the individual image depth but decreasing the overall number of visits)?}
    
    Increased exposure times decreases the Nyquist frequency and increases the probability of having trouble identifying accurate timescales of variability due to aliasing.  However, increasing the individual image depth increases the number of stars with high-cadence coverage.  For the science cases identified, total exposures longer than a minute especially risk complicating their interpretation by introducing Nyquist aliases.
    
    {\it \item What is the effect of a trade-off between uniformity in number of visits and co-added depth? Is there any benefit to real-time exposure time optimization to obtain nearly constant single-visit limiting depth?}
    
    Uniformity in exposure time is important so that variability amplitudes are not smoothed by different amounts in different images, complicating interpretation.
    
    {\it \item Are there any other potential trade-offs to consider when attempting to balance this proposal with others which may have similar but slightly different requests?}
    
    The 2--4\,hour durations of individual runs are ideal. Runs of at least 1 hour would cover timescales below the main Wide-Fast-Deep survey. If DDFs are proposed to be observed with conflicting, less rapid cadences, we suggest that at least 2--4\,hours of the $\approx 39$ hours of observations of each field should be obtained in a continuous mode.  We anticipate in particular that high-cadence observations will yield fewer detections of long-timescale transients such as supernovae, which may be part of the scientific motivation for selecting certain DDFs.
    
\end{enumerate}

\section{Performance Evaluation}

It is fairly straightforward to evaluate the performance of realizations of this DDF cadence for recording short-period variability.  We wish for the visits to achieve a high Nyquist frequency, $f_{\rm Nyq} = 1/(2\Delta t)$ where $\Delta t$ is the time between exposures, and also for the sequence durations to provide high effective frequency resolution, $\delta f= 1/T$, where $T$ is the total duration of each visit.  We will develop a MAF metric that can quantify comparatively how well different simulated survey strategies probe the short-period variability regime in the DDFs.  The product of the Nyquist frequency, the inverse of the effective frequency resolution, and signal-to-noise in the periodogram is a useful value to maximize $\propto T\sqrt{N} /\Delta t$, where $N$ is the number of observations obtained in all of the continuous observing sequences. The scientific yield for rapid stellar variables will also be proportional to the number of stars in each field with magnitudes below the typical single-visit depths, which could also factor into the decision of what percentage of the exposures of each DDF field (and the Galactic Plane mini survey) to obtain in a continuous mode.

\section{Special Data Processing}

Saving and processing the individual 15-second exposures is important for attaining a high Nyquist frequency, though this strategy will still yield incredibly valuable records of rapid photometric variables and transients if only the $2\times15$ second pairs are able to be processed.

\section{Acknowledgements}

This work was developed within the Transients and Variable Stars Science Collaboration (TVS) and the authors acknowledge the support of TVS in the preparation of this paper.

\section{References}

\begin{itemize}
\bibitem[Bellm(2014)]{ZTF} Bellm, E.\ 2014, The Third Hot-wiring the Transient Universe Workshop, 27 
\bibitem[Cortes \& Kipping(2018)]{Cortes2018} Cortes, J., \& Kipping, D.~M.\ 2018, arXiv:1810.00776 
\bibitem[Kilkenny(2007)]{Kilkenny2007} Kilkenny, D.\ 2007, Communications in Asteroseismology, 150, 234 
\bibitem[Kurtz(1990)]{Kurtz1990} Kurtz, D.~W.\ 1990, ARA\&A, 28, 607 \bibitem[Lund et al.(2018)]{Lund2018} Lund, M.~B., Pepper, J.~A., Shporer, A., \& Stassun, K.~G.\ 2018, arXiv:1809.10900 
\bibitem[Macfarlane et al.(2015)]{Macfarlane2015} Macfarlane, S.~A., Toma, R., Ramsay, G., et al.\ 2015, MNRAS, 454, 507 
\bibitem[Moffett(1974)]{Moffett1974} Moffett, T.~J.\ 1974, ApJS, 29, 1 
\bibitem[Montgomery \& Odonoghue(1999)]{Montgomery1999} Montgomery, M.~H., \& Odonoghue, D.\ 1999, Delta Scuti Star Newsletter, 13, 28 
\bibitem[Mukadam et al.(2006)]{Mukadam2006} Mukadam, A.~S., Montgomery, M.~H., Winget, D.~E., Kepler, S.~O., \& Clemens, J.~C.\ 2006, ApJ, 640, 956 
\bibitem[Pietrukowicz et al.(2017)]{Pietrukowicz2017} Pietrukowicz, P., Dziembowski, W.~A., Latour, M., et al.\ 2017, Nature Astronomy, 1, 0166 
\bibitem[Toma et al.(2016)]{Toma2016} Toma, R., Ramsay, G., Macfarlane, S., et al.\ 2016, MNRAS, 463, 1099 
\bibitem[VanderPlas \& Ivezi{\'c}(2015)]{VanderPlas2015} VanderPlas, J.~T., \& Ivezi{\'c}, {\v Z}.\ 2015, ApJ, 812, 18

\end{itemize}
\end{document}